\documentstyle[prd,aps,floats]{revtex}
\tighten
\begin{document}
\draft
\preprint{} 
\twocolumn[\hsize\textwidth\columnwidth\hsize\csname@twocolumnfalse\endcsname 
\title{Microwave Background Signals from Tangled Magnetic Fields}
\author{Kandaswamy Subramanian$^{1,2}$ and John D. Barrow$^1$ \\
$^1$Astronomy Centre, University of Sussex, Brighton BN1 9QJ, UK.\\
$^2$National Centre for Radio Astrophysics, TIFR, 
Poona University Campus, \\ 
Ganeshkhind, Pune 411 007, India. }
\maketitle

\begin{abstract}
An inhomogeneous cosmological magnetic field will create Alfv\'en-wave modes
that induce a small rotational velocity perturbation on the last scattering
surface of the microwave background radiation. The Alfv\'en-wave mode
survives Silk damping on much smaller scales than the compressional modes.
This, in combination with its rotational nature, ensures that there will be
no sharp cut-off in anisotropy on arc-minute scales. We estimate that a
magnetic field which redshifts to a present value of $3\times 10^{-9}$ Gauss
produces temperature anisotropies at the $10\mu K$ level at and below $10$
arc-min scales. A tangled magnetic field, which is large enough to influence
the formation of large scale structure is therefore potentially detectable
by future observations.
\end{abstract}

\maketitle
\date{\today}
\pacs{PACS Numbers : 98.80.Cq, 98.70.Vc, 98.80.Hw}
] \renewcommand{\thefootnote}{\arabic{footnote}} \setcounter{footnote}{0}

The origin of ordered, large-scale cosmic magnetic fields remains a
challenging problem. It is widely assumed that magnetic fields in
astronomical objects, like those in galaxies, grew by the amplification of
small seed magnetic fields by turbulent dynamo action \cite{dynam}. However,
the efficiency of the turbulent galactic dynamo is still being debated \cite
{dyndeb}. Alternatively, the galactic field could be a remnant of a
cosmological field of primordial origin \cite{primhyp}, although, as yet,
there is no entirely compelling mechanism for producing the required field.
It could be present in the initial conditions, be produced quantum
gravitationally or at a phase transition, or be generated in some way at the
end of a period of inflation \cite{euseed}. A primordial field that expanded
to contribute a present field strength of order $10^{-9}$ Gauss, tangled on
galactic scales, could also impact significantly on galaxy formation \cite
{wasser,ksjdb}. It is of considerable interest, therefore, to find different
ways of limiting or detecting such primordial fields \cite{kron}. We argue
that observations of anisotropies in the cosmic microwave background
radiation (CMBR), provide a potentially powerful constraint on such fields.
Indeed, earlier work has found that the isotropy of the CMBR already places
a limit of $6.8\times 10^{-9}(\Omega _0h^2)^{\frac 12}$ Gauss on the present
strength of any {\it uniform} (spatially homogeneous) component of the
magnetic field \cite{barrow}. ($\Omega _0$ is the present density parameter,
and $h$ the Hubble constant in units of $100kms^{-1}Mpc^{-1}$.) Here, we
obtain comparable constraints on tangled (inhomogeneous) magnetic fields as
well and highlight the distinctive fluctuation signature that they are
expected to leave in the small-scale structure of the CMBR.

We build on the results of an earlier paper \cite{ksjdb}, where we
considered the evolution and damping of cosmic magnetic fields in the early
universe. We showed that inhomogeneous magnetic fields induce rotational
Alfv\'en modes as well as compressional density and velocity perturbations
(see also ref.\cite{jed}). The compressional fluid
perturbations are of small amplitude ($\sim V_A^2<<1)$ on all scales,
because of the restoring force contributed by the pressure of the
radiation-baryon fluid. Here, $V_A$ is the Alfv\'en velocity in the
radiation era of the universe (see eqn. (\ref{alfvel}) below) and we 
set $c=1$. Compressional modes are also strongly 
damped on scales smaller than the
comoving photon diffusion length, $L_S$, by Silk damping \cite{silk}.
However, the{\it \ rotational }component of the fluid velocity, induced by
tangled magnetic fields, can be much larger (a significant fraction of $V_A$%
). Two features are important. First, the restoring force is much weaker
(since radiation-baryon pressure does not affect this mode), and
furthermore, these velocity perturbations survive damping on comoving scales
larger than $L_A\sim V_AL_S$ \cite{ksjdb}. It is these rotational velocity
perturbations, induced by the tangled field, which can produce significant
small angular scale anisotropies in the CMBR, through the Doppler effect.

We calculate this effect in two steps. First we calculate the anisotropy at
small angular scales produced by a general rotational component of the
velocity, due to the Doppler effect. Then we estimate the magnitude of this
rotational component for a given spectrum of magnetic inhomogeneities.

The equation governing the evolution of temperature perturbation $\Delta T(%
{\bf x},{\bf \gamma },\tau _0)/T_0$, can be derived from
the moments of the Boltzmann equation for photons \cite{mabert}. Here, ${\bf %
x}$ is the comoving position, $\tau $ is conformal time, $\tau _0$ its
present value, and ${\bf \gamma }$ is the direction of photon propagation.
It is conventional to expand $\Delta T/T_0$ in spherical harmonics with $%
\Delta T/T_0=\sum_{lm}a_{lm}Y_{lm}({\bf \gamma })$ and define $%
C_l=<|a_{lm}|^2>$. For rotational velocity perturbations 
a detailed calculation of $%
C_l$ has been done in Ref.\cite{vish} and \cite{huwh} in the context of
computing the Vishniac effect, and we can adopt their final result. We have
(from Eqs. (5) and (16) of \cite{huwh}), 
\begin{eqnarray}
C_l &=&4\pi \int_0^\infty {\frac{k^2dk}{2\pi ^2}}\quad {\frac{l(l+1)}2} 
\nonumber \\
\  &&\times <|\int_0^{\tau _0}d\tau g(\tau _0,\tau )V(k,\tau ){\frac{%
j_l(k(\tau _0-\tau ))}{k(\tau _0-\tau )}}|^2>.  \label{deldef}
\end{eqnarray}
Here, $V(k,\tau )$ is the Fourier transform of the rotational component of
the fluid velocity, with $k$ the co-moving wave number, and $j_l(z)$ is the
spherical Bessel function of order $l$. The 'visibility function', $g(\tau
_0,\tau ),$ determines the probability that a photon reaches us at the epoch 
$\tau _0$ if it was last scattered at the epoch $\tau $. It is given by $%
g(\tau ,\tau ^{\prime })=\dot \kappa \exp [{-\int_{\tau ^{\prime }}^\tau 
\dot \kappa (\tau ^{\prime \prime })d\tau ^{\prime \prime }}]$, where $\dot 
\kappa (\tau )=n_e(\tau )\sigma _Ta(\tau )$, $n_e$ is the electron number
density, $\sigma _T$ is the Thomson cross section, and $a(\tau )$ is the
cosmological scale factor normalised to unity at the present. We also adopt
a flat universe throughout.

For standard recombination physics, $g$ is peaked about a small range of
conformal times around the time of recombination. For the present purpose of
making approximate analytical estimates, it suffices to approximate it as a
Gaussian, and take $g(\tau _0,\tau ^{\prime })=(2\pi \sigma ^2)^{-1/2}\exp
[-(\tau ^{\prime }-\tau _{*})^2/(2\sigma ^2)]$, where $\tau _{*}$ is the
conformal epoch of ``last scattering'' and $\sigma $ measures the width of
the last scattering surface (LSS). 
Using the expressions given in Ref. \cite{husug}%
, we estimate $\tau _{*}\sim 178.2h^{-1}Mpc$ and $\sigma =11.1h^{-1}Mpc$.
Note that due to the visibility function, the dominant contributions to the
integral over $\tau $ in Eq. (\ref{deldef}) come from conformal times in a
range $\sigma $ around the epoch $\tau =\tau _{*}$. The presence of $%
j_l(k(\tau _0-\tau ))$ picks out $(k,\tau )$ values in the integrand which
have $k(\tau _0-\tau )\sim l.$

We can obtain analytic estimates of $C_l$ in two limits. First, for
wavelengths such that $k\sigma <<1$, $k(\tau _0-\tau ),$ and hence $%
j_l(k(\tau _0-\tau )),$ vary negligibly over the range of $\tau $ where $g$
is significant. So they can be evaluated at $\tau =\tau _{*}$ and
taken out of the integral over $\tau $ in Eq. (\ref{deldef}). Furthermore,
in general, $V(k,\tau )$ does not vary rapidly with conformal time within $%
\sigma $, nor does it vary rapidly with $k$ in the $k$ range around $k\sim
l/R_{*}$ where $j_l(kR_{*})$ contributes dominantly (we define $R_{*}=\tau
_0-\tau _{*})$. Thus, $V$ can also be evaluated at $\tau =\tau _{*}$ and $%
k=l/R_{*}$ and pulled out of the integrals. The remaining integral of $g$
over $\tau $ gives unity, while that over $j_l^2$ can be done analytically,
giving 
\begin{equation}
{\frac{l(l+1)C_l}{2\pi }}\approx {\frac \pi 4}\Delta _V^2(k=l{R}%
_{*}^{-1},\tau _{*}).  \label{powlar}
\end{equation}
where, $\Delta _V^2(k,\tau _{*})=k^3<|V(k,\tau _{*})|^2>/(2\pi ^2)$ is the
power per unit logarithmic interval of $k$, residing in the rotational
velocity perturbation $V$.

In the other limit, $k\sigma >>1$, for wavelengths much smaller than the
thickness of the LSS, one can treat $g$ as a
slowly-varying function compared to $j_l$ in the integral over $\tau $ in (%
\ref{deldef}). There is a cancellation due to superposition of oscillating
contributions of $j_l$ over the thickness of the LSS. An approximate
evaluation of the angular power spectrum in this case gives 
\begin{equation}
{\frac{l(l+1)C_l}{2\pi }}\approx {\frac{\sqrt{\pi }}4}{\frac{\Delta
_V^2(k,\tau _{*})}{k\sigma }}|_{k=l/R_{*}}.  \label{powsmal}
\end{equation}
Note that in this small-wavelength case, $(k\sigma >>1)$, the angular
power-spectrum is suppressed by a $1/k\sigma $ factor due to the finite
thickness of the LSS. We can now evaluate the CMBR anisotropies produced by
tangled magnetic fields.

We will assume that the tangled magnetic field is initially a Gaussian random
field. When the Hubble radius grows to encompass tangles on any particular
scale, the Lorentz force associated with the magnetic inhomogeneity induces
a fluid velocity. One would like to estimate the rotational component of
this velocity, for a general spectrum of magnetic inhomogeneities. Before
doing this however, we first consider the simpler example of the special 
{\it non-linear} cosmological Alfv\'en wave solution discussed in \cite
{ksjdb}. This will allow us to estimate in a simpler context the magnitude
of the CMBR anisotropy induced by a tangled magnetic field.

We showed in \cite{ksjdb}, that the equations of relativistic, viscous,
magnetohydrodynamics are conformally invariant during the epochs when the
radiation energy density dominates over the baryon density. We exploited
this to transform the MHD equations from the Friedmann universe to flat
spacetime, and studied a special nonlinear Alfv\'en mode solution. In this
solution, the magnetic field ${\bf B}$ is split into a uniform field, say
along the ${\bf z}$ direction, and a tangled component of {\it arbitrary
strength}, perpendicular to ${\bf z}$. All variables depend only on $z$ and $%
\tau $. In particular, ${\bf B}=[B_0{\bf z}+b_0(z,\tau ){\bf n}]/a^2$, where 
${\bf n}.{\bf z}=0$. The tangled field, $b_0(z,\tau )$, can be arbitrarily
large compared to $B_0$ and satisfies a linear damped wave equation.
Expanding the magnetic inhomogeneity in plane wave modes, as $b_0(z,\tau
)=f(k,\tau )e^{ikz}$, this is, \cite{ksjdb}, 
\begin{equation}
{\ddot f}+{\frac{\eta (\tau )k^2}{(\rho +p)a}}{\dot f}+k^2V_A^2f=0.
\label{oscill}
\end{equation}
Here, derivatives are with respect to conformal time. The associated
velocity perturbation is given by ${\bf v}(z,\tau )=v_0(k,\tau )e^{ikz}{\bf %
n;}$ from the induction equation, $v_0(k,\tau )=-i\dot f/(kB_0)$. In Eq.(\ref
{oscill}), $\eta =(4/15)\rho _\gamma l_\gamma $ is the shear viscosity
coefficient associated with the damping due to photons, whose energy density
is $\rho _\gamma $ and mean-free-path is $l_\gamma $. We have also defined
the Alfv\'en velocity $V_A$ by 
\begin{eqnarray}
V_A &=&{\frac{B_0/a^2}{(4\pi (\rho +p))^{1/2}}}={\frac{B_0}{(4\pi (\rho
_0+p_0))^{1/2}}}  \nonumber \\
\  &\approx &3.8\times 10^{-4}B_{-9}.  \label{alfvel}
\end{eqnarray}
where $\rho $ and $p$ are the energy density and pressure of the
relativistic radiation-baryon fluid; $\rho _0$ and $p_0$ their redshifted
present-day values and $B_0$ the present field strength assuming the field
is simply frozen into the uniformly expanding universe. To obtain a
numerical estimate, we have expressed $B_0$ in units of $10^{-9}$ Gauss and
taken the dominant contribution to the energy density $\rho =\rho _\gamma $,
the photon energy density, as is appropriate in the radiation era after the
epoch of $e^{+}e^{-}$ annihilation.

We can solve Eq.(\ref{oscill}) analytically in two limits. For scales (or $%
k^{-1}$), larger than the comoving Silk scale at recombination, which is $%
L_S\sim (l_\gamma (t_{*})t_{*})^{1/2}/a(t_{*})$, and assuming the fluid
starts from rest, we can neglect the frictional term compared to the driving
term in eq. (\ref{oscill}). (Here $t$ is the usual cosmic time). 
Since $\dot f(\tau _i)=0$ at some initial time,
say $\tau _i$, we have $f=f_0(k)\cos [kV_A(\tau -\tau _i)]$ and $v_0(k,\tau
)=(f_0(k)/iB_0)V_A\sin [kV_A(\tau -\tau _i)]$. For a flat universe, composed
of radiation and matter, we can integrate the Friedmann equation to obtain $%
\tau =6000h^{-1}[\sqrt{a+\Omega _\gamma }-\sqrt{\Omega _\gamma }]$ Mpc. The
phase of oscillation is then $\chi =kV_A(\tau -\tau _i)\approx
10^{-2}B_{-9}(k/(0.2hMpc^{-1}))(\tau /\tau _{*})$, where we assume $\tau
_{*}>>\tau _i$, and adopt a LSS redshift, $z_{*}\sim 1100$. Since $\chi <<1$%
, $v_0(k,\tau _{*})\approx (f_0(k)/iB_0)kV_A^2\tau _{*}\sim 0.86\times
10^{-5}(f_0(k)/iB_0)(k/0.05hMpc^{-1})(B_{-9}/3)^2$. (One can also check from
this solution that the neglect of the $\dot f$ term is a good approximation
for $kL_S<1$). The associated CMBR anisotropy will depend on the orientation
of the tangled and large-scale fields with respect to the line of sight. A
rough estimate of the maximum signal for $kL_S<<1$ is

\begin{equation}
\Delta T\sim T_0|v_0|\sim 23.3\mu K(k/0.05hMpc^{-1})(B_{-9}/3)^2;\text{ }
\label{del1}
\end{equation}
where $T_0$ is the present mean temperature of the smooth CMBR. We have
assumed that the tangled and uniform field components are of the same order.
(Also, on these scales $k\sigma <0.5,$ and there is negligible damping of
anisotropies due to the finite LSS thickness).

In the opposite limit $kL_S>>1$, damping due to photon viscosity dominates,
and one can assume the fluid reaches a terminal velocity, where $\ddot f%
\approx 0$, so $v_0\approx (if/B_0)5V_A^2/(kL_\gamma (\tau ))\sim 0.76\times
10^{-5}(if/B_0)(B_{-9}/3)^2(k/0.25h)^{-1}f_bh^{-1}$. Here, the quantity $%
L_\gamma =l_\gamma (\tau _{*})/a(\tau _{*})\sim 3.4Mpcf_b^{-1}$, is the
comoving photon mean-free-path at last scattering. We have also defined $%
f_b\equiv (\Omega _b/0.0125h^{-2})$, $\Omega _b$ being the present baryonic
density parameter (see Eq. (4.12) of Ref. \cite{ksjdb}). The oscillator is
overdamped on the relevant scales, \cite{ksjdb}. Thus, the solution with $%
\dot f(\tau _i)=0$ will have $f\sim f_0$. Further, since $L_S$ and $\sigma $
are comparable, we have $k\sigma >1$, and so there will be a $(k\sigma
)^{-1/2}$ suppression in the observed anisotropy due to the finite thickness
of the LSS. Therefore, in this case we predict that, for $kL_S>>1,$

\begin{eqnarray}
\Delta T\sim \frac{T_0c_0|v_0|}{\sqrt{k\sigma }}
\sim 12.5\mu K&&(\frac{B_{-9}}3%
)^2(\frac k{0.25hMpc^{-1}})^{-3/2} \nonumber\\
&\times& c_0h^{-1}f_b;\text{ }  \label{del2}
\end{eqnarray}
Again, we have assumed comparable tangled and homogeneous fields. The
geometrical factor $c_0<1$ incorporates the unknown orientation of the
large-scale field and the small-scale tangles, together with the precise
form of the LSS thickness suppression effects. We see that in both limits, (%
\ref{del1})-(\ref{del2}) small angular scale CMBR anisotropies $\sim 10\mu K$
could arise for magnetic fields $B_0$ $\sim 3\times 10^{-9}$ Gauss.

Let us now turn to a more general calculation of the magnetically induced
rotational-velocity perturbation, in the case when the fields 
are initially stochastic with Gaussian probability distribution. 
On galactic scales and above, the induced velocity is 
generally so small that it does not lead to
any appreciable distortion of the initial field (see above and \cite{ksjdb}%
). So, to a very good approximation, one can assume that the evolution of
the magnetic field is simply a dilution due to the uniform Hubble expansion,
with ${\bf B}({\bf x},t)={\bf b}_0({\bf x})/a^2$. (Here we assume ${\bf b}_0$
does not include any uniform component.) The Lorentz force associated
with the tangled field is then given by ${\bf F}_L=({\bf \nabla }\times {\bf %
b}_0)\times {\bf b}_0/(4\pi a^5)$. The Lorentz force still pushes on the
fluid, causing rotational velocity perturbations. We also focus on the
perturbations with co-moving length scales larger than the photon
mean-free-path at decoupling, and describe the viscous effect due to
photons, as before, in the diffusion approximation. The Fourier transform of
the linearised Euler equation for the rotational perturbations in the
baryon-photon fluid is given by 
\begin{equation}
\left( {\frac 43}\rho _\gamma +\rho _b\right) {\frac{\partial V_i}{\partial t%
}}+\left[ {\frac{\rho _b}a}{\frac{da}{dt}}+{\frac{k^2\eta }{a^2}}\right] V_i=%
{\frac{P_{ij}F_j}{4\pi a^5}}.  \label{eulerk}
\end{equation}
Here we have defined the Fourier transforms of the magnetic field, ${\bf b}%
_0({\bf x})=\sum_{{\bf k}}{\bf b}({\bf k})\exp (i{\bf k}.{\bf x})$ and the
velocity field, ${\bf v}({\bf x},t)=\sum_{{\bf k}}{\bf V}({\bf k},t)\exp (i%
{\bf k}.{\bf x})$, where ${\bf V}.{\bf k}=0$ for the transverse (vortical)
part of the velocity. Also $\rho_b$ is the baryon density. 
The fact that the Lorentz force is non-linear in ${\bf %
b}_0({\bf x})$, leads to the mode-coupling term ${\bf F}({\bf k})=\sum_{{\bf %
p}}[{\bf b}({\bf k}+{\bf p}).{\bf b}^{*}({\bf p})]{\bf p}-[{\bf k}.{\bf b}%
^{*}({\bf p})]{\bf b}({\bf k}+{\bf p})$, and the projection tensor, $P_{ij}(%
{\bf k})=[\delta _{ij}-k_ik_j/k^2]$ projects ${\bf F}$ onto its transverse
components (perpendicular to ${\bf k}$ ).

As before, we can solve Eq.(\ref{eulerk}) analytically, on scales much
larger or much smaller than the Silk scale. For $kL_s<1$, and when the fluid
starts from rest ($V_i(\tau _i)=0$), the damping due to the photon viscosity
can be neglected compared to the driving due to the Lorentz force.
Integrating the resulting first-order equation gives $V_i=G_i(\tau -\tau
_i)/(1+S_{*})$, where we have defined $G_i=P_{ij}F_j/[4\pi (\rho _0+p_0)]$
and $S_{*}=(3\rho _b/4\rho _\gamma )(\tau _{*})\sim 0.4f_b$. The $(1+S_{*})$
factor results from the reduction in the induced velocity due to baryon
inertia, an effect neglected in the non-linear Alfv\'en wave solution above.
One can also check from this solution that the neglect of viscous damping is
valid for $kL_S<1$. In the other limit, with $kL_s>>1$, we can use the
terminal-velocity approximation, neglecting the inertial terms in the Euler
equation, and simply balance the Lorentz force by friction. This leads to $%
V_i=(G_i/k)(kL_\gamma /5)^{-1}$.

To calculate the induced CMBR anisotropy, we need to define the spectrum of
the tangled magnetic field, $M(k)$. We define, $<b_i({\bf k})b_j({\bf q}%
)>=\delta _{{\bf k},{\bf q}}P_{ij}({\bf k})M(k)$, where $\delta _{{\bf k},%
{\bf q}}$ is the Kronecker delta which is non-zero only for ${\bf k}={\bf q}$%
. With this definition we also have $<{\bf b}_0^2>=2\int (dk/k)\Delta _b^2(k)
$, where $\Delta _b^2(k)=k^3M(k)/(2\pi ^2)$ is the power per logarithmic
interval in $k$ space residing in magnetic tangles, and we have changed the
summation over $k$ space to an integration. The ensemble average $<|V|^2>$ $=
$ $<V_iV_i^{*}>$, and hence the $C_l$s, can be computed in terms of the
magnetic spectrum $M(k)$. It is convenient to define a dimensionless
spectrum, say $h(k)=\Delta _b^2(k)/(B_0^2/2)$, where $B_0$ is a fiducial
constant magnetic field. We will also define the Alfv\'en velocity, $V_A$,
for this fiducial field, using Eq.(\ref{alfvel}). Finally, as a measure of
the CMBR anisotropy induced by the tangled magnetic field, we define the
quantity $\Delta T_B(l)=\{l(l+1)C_l/2\pi \}^{1/2}T_0$.

Since scales with $kL_s<1$ also generally satisfy the criterion $k\sigma <1$%
, the resulting CMBR anisotropy on these scales can be estimated using Eq.(%
\ref{powlar}). A lengthy but straightforward calculation gives, for scales
with $kL_s<1$ and $k\sigma <1$, 
\begin{eqnarray}
\Delta T_B(l) &=&T_0({\frac \pi {32}})^{1/2}I(k){\frac{kV_A^2\tau _{*}}{%
(1+S_{*})}}  \nonumber \\
\  &\approx &5.4\mu K\left( {\frac{B_{-9}}3}\right) ^2\left( {\frac l{300}}%
\right) I({\frac l{R_{*}}})  \label{largT}
\end{eqnarray}
Here, $l=kR_{*}$ as before, $S_{*}=0.4$, and we have again assumed $\tau
_{*}>>\tau _i$. 

For scales with $kL_S>1$ and $k\sigma >1$, we can use Eq.(\ref{powsmal}),
and $V_i=(G_i/k)(kL_\gamma /5)^{-1}$. In this case a similar calculation to
the one performed above gives 
\begin{eqnarray}
\Delta T_B(l) &=&T_0{\frac{\pi ^{1/4}}{\sqrt{32}}}I(k){\frac{5V_A^2}{%
kL_\gamma (\tau _{*})(k\sigma )^{1/2}}}  \nonumber \\
\  &\approx &5.6\mu K\left( {\frac{B_{-9}}3}\right) ^2\left( {\frac l{1500}}%
\right) ^{-3/2}h_{50}^{-1} I({\frac l{R_{*}}})  \label{smalT}
\end{eqnarray}
where $h_{50}=(h/0.5)$ and we have taken $f_b=1$. 
(Also, for the diffusion approximation to be
valid, $kL_\gamma (\tau _{*})<1$.)

The function $I^2(k)$ in the eqs.(\ref{largT})-(\ref{smalT}) is a
dimensionless mode-coupling integral given by 
\begin{eqnarray}
I^2(k) &=&\int_0^\infty {\frac{dq}q}\int_{-1}^1d\mu {\frac{h(q)h(|({\bf k}+%
{\bf q})|)k^3}{(k^2+q^2+2kq\mu )^{3/2}}}  \nonumber \\
&&\ \times (1-\mu ^2)\left[ 1+{\frac{(k+2q\mu )(k+q\mu )}{(k^2+q^2+2kq\mu )}}%
\right]   \label{modint}
\end{eqnarray}
where $|({\bf k}+{\bf q}|=(k^2+q^2+2kq\mu )^{1/2}$. In general, $I(k)$ can
only be evaluated numerically. We will do this elsewhere, but its order of
magnitude contribution can be estimated from the following simple case.
Suppose the magnetic spectrum has a single scale, with $h(k)=k\delta
_D(k-k_0)$, where $\delta _D(x)$ is the Dirac delta function). In that case $%
<{\bf b}_0^2>=B_0^2$ and the field points in random directions but has a
unique scale $k_0^{-1}$. The mode-coupling integral can be evaluated exactly
for such a spectrum. We find that $I(k)=(k/k_0)[1-(k/2k_0)^2]^{1/2}$, for $%
k<2k_0$, and zero for larger $k$. One can see that $I(k)$ contributes a
factor of order unity around $k\sim k_0$, with $I(k_0)=\sqrt{3}/2$. For more
complicated magnetic spectrum, with a multitude of scales, $I(k)$ can be
thought of as a superposition of these elementary contributions, and could
be somewhat larger.

From Eqs.(\ref{largT})-(\ref{smalT}), we see that in general for a tangled
field of order $B_0\sim 3\times 10^{-9}G$, one expects a RMS CMBR anisotropy
of order $5\mu K$ or larger, depending on the contribution of $I(k)$ and the
value of $l$. The anisotropy in hot or cold spots could be several times
larger, because the non-linear dependence of $C_l$ on $M(k)$ will imply a
non-Gaussian statistics for the anisotropies. It should also be emphasised
that in standard models the $C_ls$ have a sharp cut off for $l>R_{*}/L_S$,
due to Silk damping. But strong damping of Alfv\'enic perturbations is
expected only on scales smaller than $V_AL_S$ \cite{ksjdb}. Moreover, for
small scale rotational perturbations, the damping due to the finite
thickness of the LSS is also milder than for compressional modes. 
We note in passing that such rotational perturbations could also induce
a ``magnetic'' type polarisation anisotropy. We will
estimate this effect elsewhere.

Several satellite and interferometric experiments are planned which will map
the small angular scale anisotropies at the levels that these calculations
predict. We have identified a new physical effect which can produce a
detectable signal in the microwave background on arc minute scales. If
magnetic fields play a role in the development of large-scale structure in
the Universe then we should be able to detect the effects of tangled
primordial magnetic fields.

{\it Acknowledgments:} The authors were both supported by PPARC. KS also
thanks Simon White and the Max-Planck Institute for Astrophysics, Garching,
Rainer Beck, Richard Wielebinski and the Max-Planck Institut for Radio
Astronomy, Bonn for hospitality during the course of this work, and
T. R. Seshadri for a useful discussion.

\end{document}